\documentclass[letterpaper]{article} % DO NOT CHANGE THIS
\usepackage{aaai23}  % DO NOT CHANGE THIS
\usepackage{times}  % DO NOT CHANGE THIS
\usepackage{helvet}  % DO NOT CHANGE THIS
\usepackage{courier}  % DO NOT CHANGE THIS
\usepackage[hyphens]{url}  % DO NOT CHANGE THIS
\usepackage{graphicx} % DO NOT CHANGE THIS
\usepackage{natbib}  % DO NOT CHANGE THIS AND DO NOT ADD ANY OPTIONS TO IT
\usepackage{caption} % DO NOT CHANGE THIS AND DO NOT ADD ANY OPTIONS TO IT
\usepackage{algorithm}
\usepackage{algorithmic}

\usepackage{newfloat}
\usepackage{listings}
\DeclareCaptionStyle{ruled}{labelfont=normalfont,labelsep=colon,strut=off} % DO NOT CHANGE THIS
\floatstyle{ruled}
\newfloat{listing}{tb}{lst}{}
\floatname{listing}{Listing}
\usepackage{todonotes}
\usepackage{helvet,mathpazo}
\usepackage{microtype}
\usepackage{textpos}
\usepackage{amsmath,siunitx}
\usepackage{booktabs,colortbl,xcolor}
\usepackage{graphicx,wrapfig}
\usepackage{enumitem}
\usepackage{subfigure}
\usepackage{multirow}

\title{Leveraging Vision-Language Models for Granular Market Change Prediction}

\author{
    Christopher Wimmer\textsuperscript{\rm 1}, Navid Rekabsaz\textsuperscript{\rm 1 \rm 2}
}

\affiliations {
\textsuperscript{\rm 1}Johannes Kepler University Linz\\
\textsuperscript{\rm 2}Linz Institute of Technology, AI Lab, \\
cedwimmer@gmx.at,
navid.rekabsaz@jku.at\\
}

\usepackage{bibentry}
\begin{document}

\maketitle

\begin{abstract}
Predicting future direction of stock markets using the historical data has been a fundamental component in financial forecasting. This historical data contains the information of a stock in each specific time span, such as the opening, closing, lowest, and highest price. Leveraging this data, the future direction of the market is commonly predicted using various time-series models such as Long-Short Term Memory networks. This work proposes modeling and predicting market movements with a fundamentally new approach, namely by utilizing image and byte-based number representation of the stock data processed with the recently introduced Vision-Language models. We conduct a large set of experiments on the hourly stock data of the German share index and evaluate various architectures on stock price prediction using historical stock data. We conduct a comprehensive evaluation of the results with various metrics to accurately depict the actual performance of various approaches. Our evaluation results show that our novel approach based on representation of stock data as text (bytes) and image significantly outperforms strong deep learning-based baselines.

\end{abstract}

\section{Introduction} \label{Introduction_MAIN}
% !TeX encoding = UTF-8
% !TeX root = MAIN.tex

An experienced (human) investor in the stock market develops the ability to visually interpret and read the price trend, traditionally done using a line or candlestick chart. Based on this information, investor form an overall picture and make a decision either to buy or sell a stock. The rapid development of deep learning models in recent years has suggested new frontiers to the decades-long research for the ways to predict the future movement of the stock market. In this approach, a deep learning model looks at the technical data -- the historical values of price chart -- in order to draw conclusions from the past movements for the future changes.

The recent advancements in Vision-Language models, e.g. in the CLIP~\cite{CLIP} model, offer a new set of possibilities for processing data in parallel in various modalities. The training of these models allows the network to learn visual concepts through natural language supervision, where in the case of CLIP, the model is trained on a dataset of 400 million image-text pair. In this work, we study the benefits of leveraging these capabilities for market prediction, by using both the image and text pipeline of CLIP to extract meaningful features from the technical data of the German share index. Concretely, the contributions of this work are as follows:
\begin{itemize}
\item We build a framework for data gathering and assessment of market forecasting based on the German share index on an hourly basis.
\item We provide a unique strategy for forecasting the market using CLIP based on the numeric and picture representations of the data, accompanied with Long Short Term Memory (LSTM) \cite{hochreiter1997longLong_short_term_memory} model.
\item We conduct extensive experiments showing considerable improvements in market prediction using the suggested CLIP-based technique when compared with state of the art baselines.
\item We examine the hypothetical case of making decisions faster than other traders, characterized by assigning two distinct labeling schema. 
\item Using an interpretation tool, we analyze the predictions to assess possible similarities of the image-based forecasting with human decision-making processes.
\end{itemize}

Overall, our study shows the remarkable performance of the CLIP approaches and demonstrates that our suggested method to extract information can outperform the trader's speed.

The work is organized as follows: Section~\ref{Related Work} discusses the background and related work in. The dataset, the labeling techniques, and the proposed methods are presented in Section~\ref{Methodology_MAIN}. The details of the experiments and the evaluation metrics are provided in Section~\ref{Experiment Design_MAIN}, and the results and a short excursion into the interpretability of CLIP are discussed in Section~\ref{Results and Interpretability_MAIN}. Finally, the work is concluded in Section~\ref{Conclusion_MAIN}, and the  followed by the appendix.

\section{Related Work}
\label{Related Work}
% !TeX encoding = UTF-8
% !TeX root = MAIN.tex

\begin{figure*}[t]
    \centering

\subfigure[LSTM]{\includegraphics[width=0.25\textwidth]{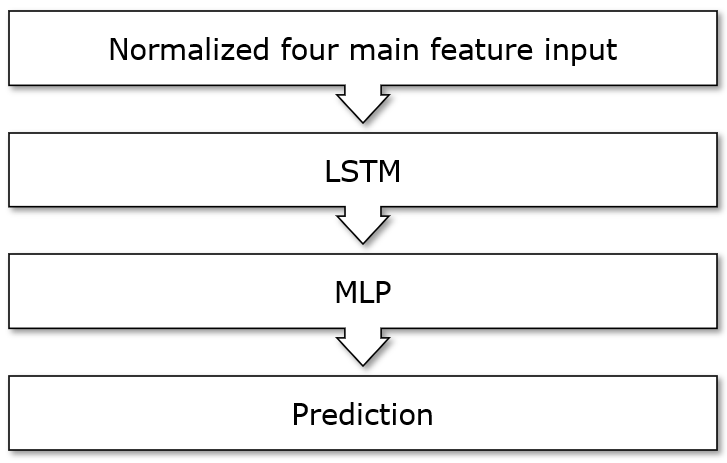}\label{fig:thesis/images/0_4/LSTM_Workflow.png}}
~
\subfigure[Stacked-LSTM]{\includegraphics[width=0.25\textwidth]{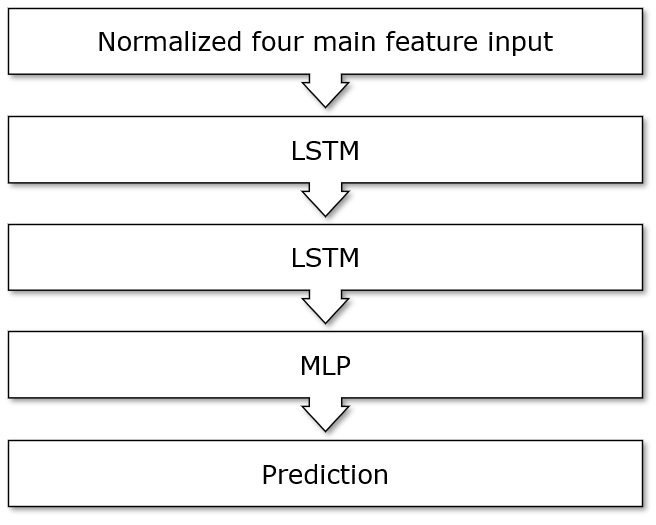}\label{fig:thesis/images/0_4/Stacked_LSTM_Workflow.png}}
~
\subfigure[DAIN-LSTM]{\includegraphics[width=0.25\textwidth]{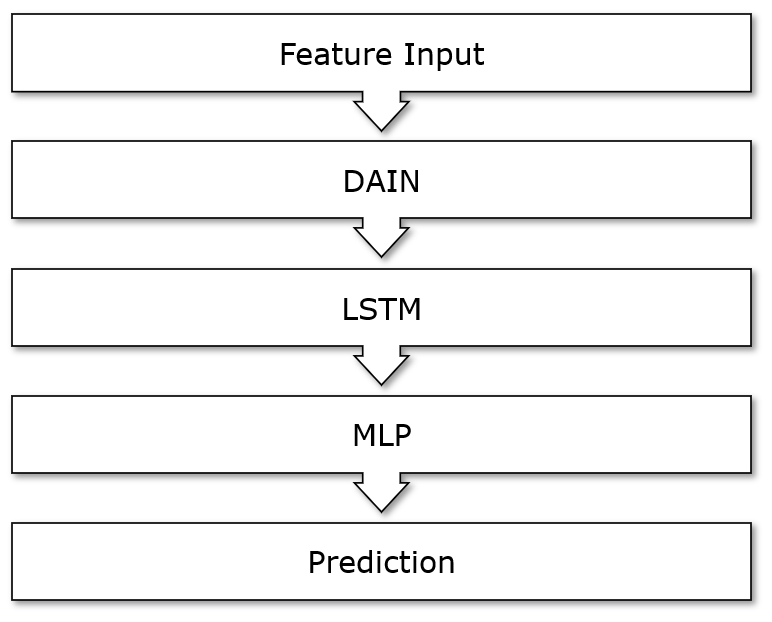}\label{fig:thesis/images/0_4/DAIN_LSTM_Workflow.png}}

\subfigure[CLIP-LSTM (technical text)]{\includegraphics[width=0.25\textwidth]{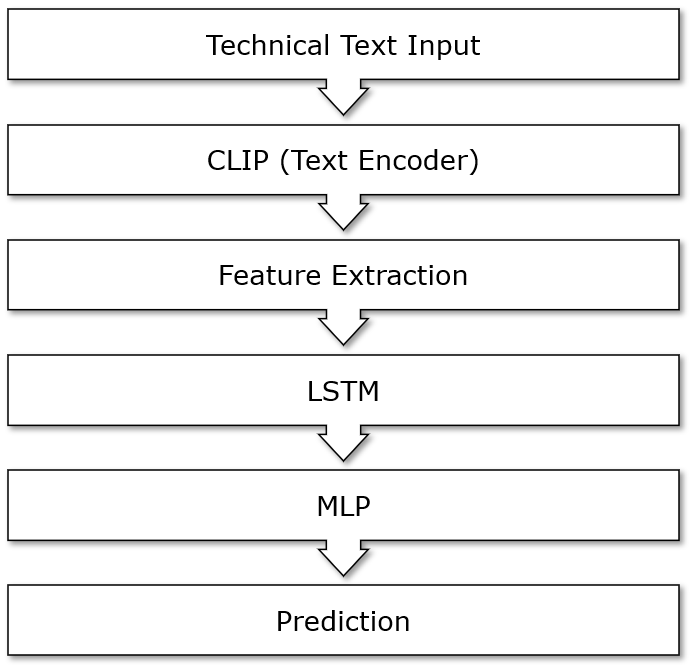}\label{fig:thesis/images/0_4/CLIP_LSTM_TEXT_Workflow.png}}
~
\subfigure[CLIP-LSTM  (technical image)]{\includegraphics[width=0.25\textwidth]{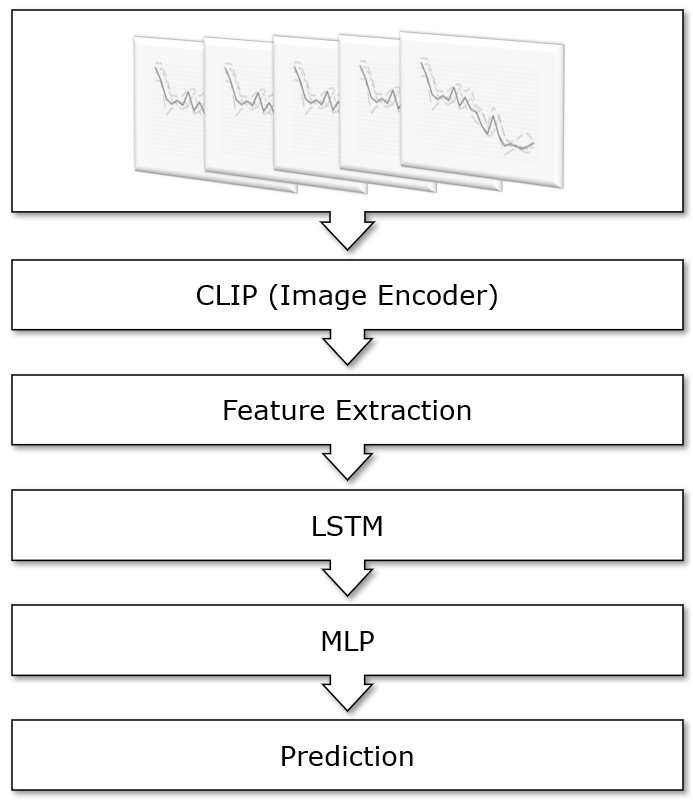}
\label{fig:thesis/images/0_4/CLIP_LSTM_PICTURE_Workflow.png}}

\caption{Schematic representation of various approaches.}
\label{fig:models}
\end{figure*}

Financial sectors are increasingly using deep-learning and machine learning technology in a variety of ways as discussed in \citet{15PredictingStockMarketTrends}. In this section, we review a variety of predicting techniques for stock prices. For a detailed background on the technical terminologies in the finance domain used in this paper, please refer to Appendix~\ref{sec:appendix:background-finance}.

%\subsection{Technical Data Approaches} \label{Technical Data Approaches_Related Work}
A large body of work exclusively use technical market data, namely the chart data of different stock prices. As representative studies in this domain, \citet{CNNBiLSTMAMmethodforstockpriceprediction} investigate the feasibility of employing Convolutional Neural Networks (CNN)~\cite{lecun1995convolutional} in combination with the LSTM architectures to forecast the following day's closing price using stock information. \citet{AEIDNET} forecast ten well-known stocks and Stock Technical Indicators over ten years using a dense neural network in combination with an autoencoder in the short, medium, and long term. \citet{Convolutionalneuralnetworkforstocktradingusingtechnicalindicators} analyzes publicly accessible data on stocks, where ten technical indicators are retrieved as feature vectors from this data and are then transformed to images and supplied into a CNN model. Stock closing prices are manually classified as sell, buy, or hold and the evaluation is measured with F1 score.

Several studies utilize LSTM for market prediction. \citet{LSTMbaseddecisionsupportsystemforswingtradinginstockmarket} provide a decision support system using an LSTM architecture to create a report that includes the expected values of a company's stock for the next 30 days. \citet{StockMarketPriceMovementPredictionWithLSTM} show that a high accuracy rate for predicting the future movement of a stock may be achieved by the use of historical chart data in conjunction with an LSTM. Such an approach is considered in our study as a baseline method. Further on, \citet{ADeepLearningModelforPredictingBuyandSell} compare the results of the LSTM algorithm to those of classical machine learning algorithms such as Support Vector Machine, Multi-layer Perceptron, decision tree, random forest, logistic regression and $k$-nearest neighbors. Similar to the study in hand, they conduct the experiments on high, low, opening, and closing prices of the stock chart as input to predict buy/sell signal. They conclude that among the mentioned algorithms, LSTM performs the best. Our work contributes to this line of research by proposing a novel prediction method using Vision-Language models.

%\citet{Forecastingtheovernightreturndirectionofstockmarketindexcombiningglobalmarketindices} suggest that anticipating index performance overnight (close-to-open) is becoming more prevalent, and that global markets are interconnected. They use convolution units to extract information from each branch input layer representing global stock market indexes in order to predict the overnight return direction of a specific stock market index. The factors are then combined in order to determine the day direction of the night return.

\begin{table*}[t]
%\scriptsize
    \centering
    \begin{tabular}{cccccc}
    \toprule
    Date and Time &    Close &     Open &     High &      Low  &  Label \\\midrule
    2020-04-21 22:00:00 &  10298.7 &  10282.5 &  10306.1 &  10277.4 &      1 \\
    2020-04-21 23:00:00 &  10316.0 &  10301.4 &  10320.8 &  10300.5  &      0 \\
    ... &  ... &  ... &  ... &  ... &   ... \\
    \end{tabular}

    \vspace{3mm}
    \begin{tabular}{c}
    %\toprule
    Equivalent Input Text \\
    \midrule
    ``Date:21/04/2020 Time:22 Close:10298.7 Open:10282.5 High:10306.1 Low:10277.4'' \\
    ``Date:21/04/2020 Time:23 Close:10316.0 Open:10301.4 High:10320.8 Low:10300.5'' \\
    ...  \\
    %\bottomrule
    \end{tabular}

    \caption{Top: an exemplary extract from the technical German share index dataset. Bottom: the equivalent concatenated text used in our CLIP-based approach.}
    \label{tbl:technicalData}
    
\end{table*}

Another line of research uses various other resources such as news, Twitter data, company calls to bring in extra signals for prediction. As examples of studies, \citet{TransformerbasedDeepIntelligentContextualEmbeddingforTwittersentimentanalysis} mine online Twitter data and use Twitter sentiment analysis to improve the decision-making processes of stock market prediction. \citet{araci2019finbertFinBERT_Financial_Sentiment_Analysis_with_Pre_trained_Language_Models} propose FinBert, a BERT-like \cite{devlin2019bertBERT_Pre_training_of_Deep_Bidirectional_Transformers_for_Language_Understanding} Transformer architecture trained on financial data, and use the model to classify the sentiment of financial language. \citet{PortfolioOptimizationwith2DRelativeAttentionalGatedTransformer} observe a favorable correlation between the rise of the stock market and the level of positive sentiment on Twitter. That is, despite the high volume of hopeful tweets, the markets were more likely to rise. To forecast future stock market movement in China, \citet{Exploitinginvestorssocialnetwork} creates a fundamental dataset using text data from a Chinese social networking site. This data is used to estimate the stock market's future movement in China. The data is expressed in stock-related words in order to maximize the influence of the text data on the stock market.

As hybrid approaches using technical and textual resources, \citet{Rekabsaz_2017} predict volatility by analyzing annual filings of companies for forecasting market volatility. In their work, the filings sentiment is inferred by leveraging a novel word embedding-based term weighting models~\cite{10.1145/2983323.2983833} where the similarity thresholds are set on basis of the uncertainty of embedding models~\cite{10.1145/3404835.3462951,10.1007/978-3-319-56608-5_31}. The word embedding methodology surpasses previous approaches. More recently, \citet{S_I_LSTM} look at stock prices, technical indicators, and non-traditional data sources such as stock postings and financial news. To develop an investor sentiment index, fundamental text data is analyzed using a CNN pipeline, while technical data is handled using an LSTM architecture. %\citet{HISASMFM} make predictions about the future motion of the price of several stocks based on a variety of fundamental variables and past price data from the chart. After processing the different kinds of data sources they use a Long Short Term Memory (LSTM) architecture to get a prediction.

\section{Methodology} 
\label{Methodology_MAIN}

% \begin{figure}[t]
%     \centering
%     \includegraphics[width=0.5\textwidth]{thesis/images/0_4/normalisation_problem.png}
%     \caption{Illustration of the possible occurrence of moving out of the in the training set defined window.}\label{fig:thesis/images/0_4/normalisation_problem.png}
% \end{figure}

\begin{figure}[t!]
    \centering
    \includegraphics[width=0.8\columnwidth]{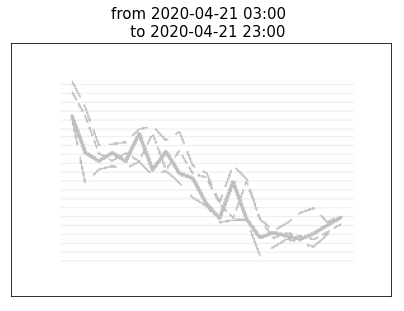}
    %\qquad
    \caption{The equivalent image of the exemplary extract from the technical German share index dataset shown at the top of Table~\ref{tbl:technicalData}.}
    \label{fig:technicalData}
\end{figure}

Since stock market is in the form of a time series, we first need to transform the content of input data into a meaningful features. Such features can then be given to a time-series prediction model such as LSTM to predict for the stock price direction. In this section, we first discuss some strong baseline methods for market prediction using technical data, followed by explaining our approach to creating these features and forecasting market movements. Figure~\ref{fig:models} demonstrates a schematic view of the models discussed in following. In Appendix~\ref{sec:appendix:background-deep}, we provide the background on the general principles of the LSTM, Deep Adaptive Input Normalization \cite{passalis2019deepDeep_Adaptive_Input_Normalization_for_Time_Series_Forecasting} (DAIN), and CLIP components, used in the following market prediction models.

\subsection{Baseline Market Prediction Models using Technical Data}
\paragraph{LSTM} is the first baseline model and uses a basic LSTM component shown in Figure~\ref{fig:thesis/images/0_4/LSTM_Workflow.png}. The model receives four features, namely open, high, low, and close prices. The price values are first normalized using: $x_{norm} = \frac{x - \mu_{train} }{\sigma_{train}}$, where $x_{norm}$ is the normalized data point, $x$ denotes the original data points, $\mu_{train}$ is the mean of the training samples, and $\sigma_{train}$  is the standard deviation of the training samples. We use the train set to compute the mean and standard deviation of each feature and normalize the input vectors. To predict market direction, the model utilizes the normalized four features as input to the LSTM network, with the final hidden state becoming the input to a multi-layer perceptron after the sequence is processed. This output after applying a softmax layer is used to predict short/long trade. The loss during training is determined by binary cross entropy. 

%When the price trends cross above or below the estimated normalization parameter's price range a normalization issue emerges (windowing problem). The normalization technique is still legitimate but the information used for normalization must come exclusively from the train set, as otherwise, knowledge about the future course will flow into the data, which is then fed into the LSTM in the following phase. This is expressly prohibited and is thus treated in the manner stated.

\paragraph{Stacked-LSTM} expands the basic LSTM baseline with a stacked LSTM architecture, in which the hidden state of one LSTM layer is used as the input to the next LSTM layer as shown in Figure~\ref{fig:thesis/images/0_4/Stacked_LSTM_Workflow.png}.

\paragraph{DAIN-LSTM} to better approach the price normalization issue, our next baseline uses DAIN, a novel deep-learning component which separately learns the normalizing parameters. As shown in Figure~\ref{fig:thesis/images/0_4/DAIN_LSTM_Workflow.png}, the resulting output of the DAIN is fed into the LSTM component, which has the same purpose of forecasting a positive or negative change.

\subsection{Our Vision-Language Approach}
The CLIP model \cite{CLIP} is capable of understanding text and recognizing what is shown on an image. In the following, we explain our two market prediction models based on the text and image representation of the technical data.   

\paragraph{CLIP-LSTM (technical text)} first turns each existing data point to a sentence by concatenating the labels and the corresponding values. As an example, the top of Table~\ref{tbl:technicalData} shows a data point for a specific date and hours which is used in the previous baseline models. In this CLIP-based model, we first reformat each data as a text sequence shown in Table~\ref{tbl:technicalData} bottom. Next, we use the CLIP pretrained text encoder model to create feature vectors from the given text inputs. The feature vectors are given as inputs to the LSTM network to make a final forecast of the course direction in the future. The model is depicted in Figure~\ref{fig:thesis/images/0_4/CLIP_LSTM_TEXT_Workflow.png}.

\paragraph{CLIP-LSTM (technical image)} first creates a 512-dimensional representation of the data in the form of a line chart referred to as technical image. This line chart is a common format of representing the data of a stock as explained in more detail in Appendix A. To help the network differentiate between the four primary components, the appropriate lines are shown with varying line widths and line types. To show the height of the movement, also known as volatility, horizontal thin gray lines at a distance of 20 pip are used in each instance. Each produced technical image has a matching label that indicates the sign (positive sign is label 1, negative sign is label 0) of the delta value between the closing price of some hours in the future and the last shown date (on hourly basis) on the line chart. 

These line chart images are fed into the CLIP image encoder, and the vectors created by the CLIP architecture's final pooling layer are given to the LSTM model to calculate a final prediction. The workflow is depicted in Figure~\ref{fig:thesis/images/0_4/CLIP_LSTM_PICTURE_Workflow.png}. 

%\paragraph{CLIP-LSTM (technical image and technical text)} \label{CLIP-LSTM (technical image and technical text)_methodology} We combine the two distinct approaches of leveraging CLIP as a backbone for feature generation in this approach. On the one hand, the image encoder is used to process the line chart pictures, while the CLIP text encoder is used to process the technical texts. We perform a concatenation of the generated feature vectors and a transmission to the following LSTM architecture, which after six hours learns a prediction for the sign of the delta difference. Figure~\ref{fig:thesis/images/0_4/CLIP_LSTM_PICTURE_AND_TEXT_Workflow.png} depicts the experiment's setup.

\section{Experiment Design} 
\label{Experiment Design_MAIN}
This section explains the details of experiments, namely the dataset, models, training and hyperparameters, and the evaluation metrics.

\subsection{Dataset}
%In this chapter we discuss the kind of data we utilize and the technique we use in defining the labels throughout the remainder of the thesis.

\begin{figure}[t]
    \centering
    \includegraphics[width=\columnwidth]{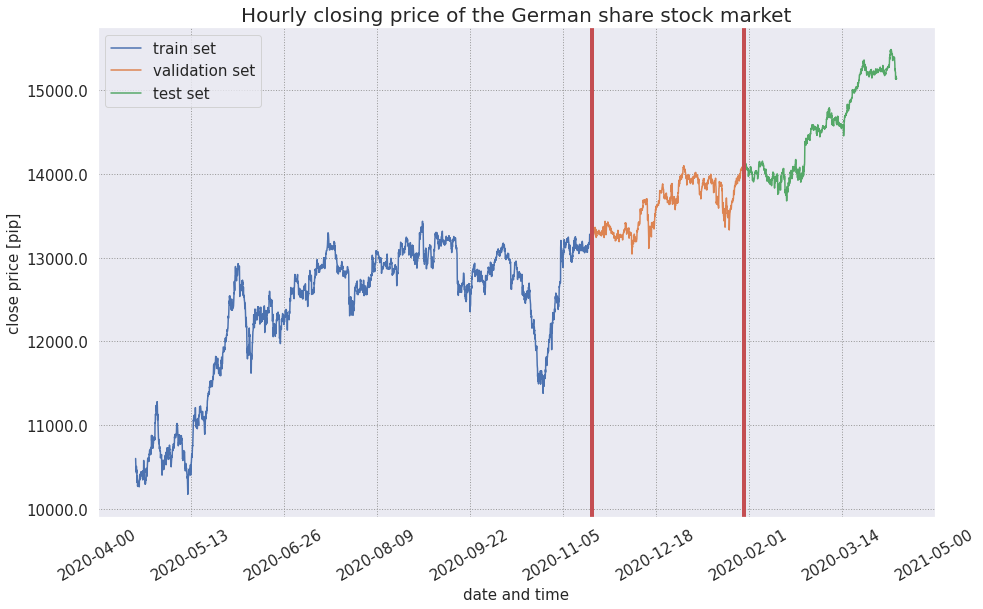}
    \caption{Schematic representation of the division of the train set, the validation set and the test set.}
    \label{fig:thesis/images/0_3/time_set.png}
\end{figure}

We collect hourly technical data of the German share index from the ``IG Markets'' platform in the standard format of the four major values in each hour. Table \ref{tbl:technicalData} shows an example of this data. Our dataset covers the time span from 2020-04-21 02:00:00 to 2021-04-20 23:00:00. We divide the data set into three subsets of train, validation and test. The first 60\% of the dataset is utilized for training, the next 20\% for validation, and the remaining 20\% for testing as shown in Figure~\ref{fig:thesis/images/0_3/time_set.png}. Table~\ref{tbl:datasets_statistics} reports the statistics of the subsets. For each data point, we define two possible target labels explained below.

%Due to the fact that this data has not been purchased, it is therefore unsuitable for commercial use and will not be published. Personal access to the brokers website, is free, however the time-consuming procedure of storing the data through the provider's source page should be addressed.

%The percentage increase in positive deltas and the growing segregation of mean deltas provide an illustration of the price growth of the various sets depicted in Figure~\ref{fig:thesis/images/0_3/time_set.png}.

%Looking at the figure, we can see that the train set contains strong volatilities in both directions to guarantee that the networks perceive and learn from all types of movement during training. The validation set is a fairly calm and stable period of time. This is used to pick the multiple trained networks and so presupposes the network's ability to generalize. The test set is used to determine the real performance of the networks following their selection based on the validation set and the consequent real performance of the networks. 
% It should be emphasized that the test set is inherently biased toward the long side, making the precision rate of the short trades critical.

\begin{table}[t]
\scriptsize
\centering
\begin{tabular}{lccc}
\toprule
{} &         Train set &  Validation set &  Test set \\
\midrule
Positive deltas [\%]           &      \textcolor{white}{0-}53.4 &           \textcolor{white}{0-}55.1 &     \textcolor{white}{0-}56.3 \\
Maximum positive delta [pip] &      \textcolor{white}{-}491.3 &           \textcolor{white}{-}317.7 &     \textcolor{white}{-}343.3 \\
Maximum negative delta [pip] &                         -408.1 &          -356.6 &    -241.4 \\
Mean delta [pip]             &       \textcolor{white}{00-}3.0 &            \textcolor{white}{00-}3.6 &      \textcolor{white}{00-}5.0 \\
Std delta [pip]              &      \textcolor{white}{-1}91.0 &            \textcolor{white}{-1}68.3 &     \textcolor{white}{-1}56.2 \\
\bottomrule
\end{tabular}
\caption{Statistics of the technical data sets}
\label{tbl:datasets_statistics}
\end{table}

\paragraph{Standard Label} The standard target label in this work is defined as the sign of the delta, calculated between the current closing price and the closing price six hours in the future. Six hours is chosen based on the assumption that it is a standard time period for the intraday stock market, as various markets such as Europe, Asia, and America open at different times due to the time difference. A market value independent forecast can be achieved by predicting the direction (positive or negative sign of the delta in six hours) of the market. Hence, Standard Label is equal to 1, if close price at time step $t$ is less than the close price at time step $t + 6$, and 0 otherwise. This standard label is shown in the examplory data in Table~\ref{tbl:technicalData}. 

\paragraph{Delayed Label} Given that all market participants have access to the same information on the present and historical price of a chart, the obvious conclusion is that the trader who first accurately analyzes the market's development using the available data may gain advantage. This implies that a system must rapidly evaluate and interpret historical data. To validate this assumption, we define the Delayed Label as the sign of the delta value in the span of one to seven hours (one hour delay). Hence, Delayed Label is equal to 1 if close price at time step $t+1$ is less than the close price at time step $t + 7$, and 0 otherwise.

\subsection{Models and Training} 
The specifications of the models are explained in the following. For each of the models, we perform an exhaustive hyperparameter random grid search. Table~\ref{hyperparameter_tab} in Appendix~\ref{sec:appendix:experiment} reports the resulting hyperparameters, which performed the best in terms of the F1 Score on the validation set. Detailed information of on the training procedure of the models is provided in Appendix~\ref{sec:appendix:experiment}.

\paragraph{Random} 
We execute one trade every hour in a random direction (long or short). All transactions are closed after six hours, identical to the other algorithms.

\paragraph{Always Long} We design this algorithm in such a way that it executes a long trade every hour, implying that the market is continually rising. This is to exploit the positive bias of the price trend.

\paragraph{Always Short} We define the inverse approach of the Always Long algorithm, which executes a short transaction every hour and exits it six hours later.

\paragraph{LSTM} We train two sequences with varying lengths to provide a fair comparison with other networks. The model "LSTM" in the results refers to a 24-hour sequence, whereas the designation "LSTM (long sequence)" refers to a 48-hour input sequence.

\paragraph{Stacked-LSTM} We use a 48-hour input sequence, since the standard LSTM architecture performs better with a longer period.

\paragraph{DAIN} We feed the input features without normalization into the DAIN approach with a 48-hour input sequence. The resulting output of the DAIN is the given to the LSTM component, which forecasts the positive/negative changes in six hours.

\paragraph{CLIP-LSTM (technical image)} For creating the line charts, we select a time frame of 20 hours with four major values assigned to each hour. We feed a five-image input sequence (time dependent with shift of one hour) into the CLIP image encoder part and give the resultant embeddings into the LSTM architecture to get the prediction of the label. The CLIP version "clip-vit-base-patch32" is used for all CLIP-based approaches.

%To assess the effect of pretraining of the CLIP model and therefore the necessary pretraining of CLIP, the same approach as described above will be used as a baseline, with the exception that the CLIP network is initialized randomly. Thus, although the network's size remains constant, the actual performance, as shown later, is generated by the time-consuming process of combining large batches of images with their respective corresponding describing texts. 

\paragraph{CLIP-LSTM (technical text)} We generate an text string for every hour that is then given to the CLIP text encoder in a 24-hour input sequence fashion. The calculated embeddings are used in combination with the LSTM architecture to predict the future direction of the stock in six hours. %To further emphasize the critical need of the powerful training procedure stated in \cite{CLIP}, a second, identically constructed attempt is created in this approach, where the CLIP text encoder will be initialized randomly and should again serves as a baseline.

%\paragraph{CLIP-LSTM (technical image and technical text)} The combination of a 24-hour technical text sequence and a five technical image input sequence sequence is delivered to the appropriate CLIP encoders, ensuring that the end date of each time series is identical. We next transmit the generated data to the LSTM architecture, which computes a six-hour prediction.

\begin{table*}[t]
\centering
\begin{tabular}{lccccccc}
\toprule
\multirow{2}{*}{\textbf{Models}} &  \multirow{2}{*}{\textbf{F1}} &  \multirow{2}{*}{\textbf{MCC}} &  \textbf{Balanced} & \textbf{Precision}     & \textbf{Precision}  & \textbf{Pip Balance}     & \textbf{Pip Balance} \\
&      &      &  \textbf{ACC} & \textbf{(Short)} & \textbf{(Long)} & \textbf{(Short)} & \textbf{(Long)} \\
\midrule
\multicolumn{8}{c}{\textbf{Standard Label}}\\
\midrule

Random                          & 0.52  & -0.09                      & 0.45  &   36.91  &  54.01  & -1049.19                            & -115.69  \\
Always Long                     & 0.73  &  0.00  & 0.50  &   0.00  &  58.03  & 0.00        &  933.49 \\
Always Short                    & 0.00  &  0.00  & 0.50  &   41.96  &  0.00  &  -933.49         &  0.00 \\
LSTM & 0.39  & -0.02  & 0.49  &   43.36  &  54.15 & -4461.57 & 1390.97  \\
LSTM (long sequence) & 0.03  &  \textcolor{white}{-}0.01  & 0.50  &   43.53  &  \textbf{58.59}  &  -5671.50 &    \textcolor{white}{0}492.20  \\
Stacked-LSTM (long sequence)    & 0.20  & -0.02  & 0.49  &   43.06  &  54.39  & -5441.57 &   \textcolor{white}{0}722.13   \\
DAIN-LSTM & 0.50  & -0.01 & 0.49  &   42.94  &  56.06  &  -3495.80 &  2516.40   \\
%Random CLIP-LSTM (image)        & \textbf{0.70}  & -0.01  & 0.50  &   41.35  &  56.11  \\
CLIP-LSTM (image)               & 0.65  & \textcolor{white}{-}\textbf{0.06}  & \textbf{0.53}  &   49.10  &  57.82 &   \textcolor{white}{0}-346.13 &  4878.77 \\
%Random CLIP-LSTM (text)         & 0.58  & -0.02  & 0.49  &   42.22  &  55.33  \\
CLIP-LSTM (text) & \textbf{0.70}  &  \textcolor{white}{-}\textbf{0.06}  & 0.52  &   \textbf{52.48}  &  56.98  &  \textcolor{white}{-}\textbf{1409.97}  &  \textbf{6634.87} \\
%CLIP-LSTM (image and text)      & 0.61  &  \textcolor{white}{-}0.05  & 0.52  &   47.23  &  57.63  \\

\midrule
\multicolumn{8}{c}{\textbf{Delayed Label}}\\
\midrule
Random                          & 0.51  &  0.01  & 0.51  &   46.27  &  55.17  & -387.80     & 843.00  \\
Always Long                     & 0.70  &  0.00  & 0.50  &   0.00  &  54.40  & 0.00        &  1230.80 \\
Always Short                    & 0.00  &  0.00  & 0.50  &   45.60  &  0.00  &  -1230.80         &  0.00 \\

LSTM & 0.23 & -0.05 & 0.48 &   42.46 &  50.76 &    -5557.17 &   \textcolor{white}{0}\textcolor{white}{-}503.63 \\
LSTM (long sequence) & 0.18 & \textcolor{white}{-}0.00 & 0.50  &   43.57 &  55.93 &  -5244.63 &  \textcolor{white}{-}1052.67\\
Stacked-LSTM (long sequence) & 0.27 & -0.03  & 0.49  &   42.23 &  53.59 & -4597.43 & \textcolor{white}{-}1884.50  \\
DAIN-LSTM & 0.49 & -0.02  & 0.49  &   42.63 &  55.52  &  -3845.63 &  \textcolor{white}{-}2287.67 \\
%Random CLIP-LSTM (image)          & 0.70 & -0.03 & 0.49  &   25.64 &  55.90 \\
CLIP-LSTM (image) & 0.66 & \textcolor{white}{-}\textbf{0.07} & 0.53 &   50.14 &  57.93  &    \textcolor{white}{0}\textcolor{white}{-}\textbf{351.37} &  \textcolor{white}{-}\textbf{5617.77} \\
%Random CLIP-LSTM (text)           & 0.51 & -0.01 & 0.50 &   43.48 &  55.66 \\
CLIP-LSTM (text) & \textbf{0.71} & \textcolor{white}{-}0.05 & 0.51 &   \textbf{53.30} &  56.70 &     \textcolor{white}{0}\textcolor{white}{0}\textcolor{white}{-}15.63 &  \textcolor{white}{-}5282.03 \\
%CLIP-LSTM (image and text) & 0.61 & \textcolor{white}{-}0.06 & \textbf{0.52} &   47.96 &  \textbf{58.34} \\
\bottomrule
\end{tabular}
\caption{Mean test set results of the top three trained models (chooses by the highest F1 Score in the validation set) with the Standard Label approach. Due to space limitations, "image" is used for "technical image", "text" is used for "technical text" and "image and text" is uses for "technical image and technical text".}
\label{tab:Scores_table_normal}
\end{table*}

\begin{figure*}[t]
    \centering
\subfigure[True-long]{\includegraphics[width=0.55\textwidth]{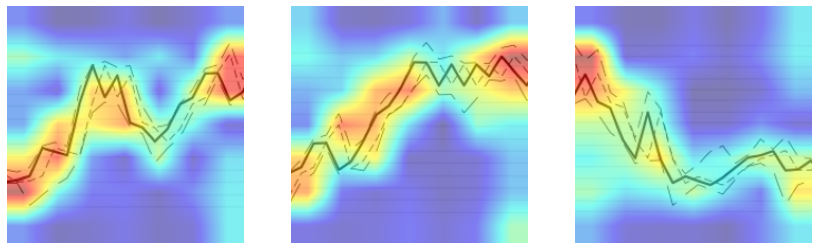}\label{fig:thesis/images/0_6/true_long_interpret.png}}

\subfigure[True-short]{\includegraphics[width=0.55\textwidth]{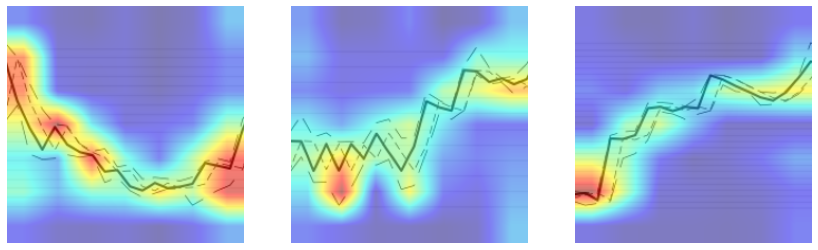}
\label{fig:thesis/images/0_6/true_short_interpret.png}}
\caption{Interpretability by the CLIP network.}
\end{figure*}

\begin{figure*}[t!]
    \centering
    \subfigure[CLIP-LSTM (image)]{\includegraphics[width=0.98\columnwidth]{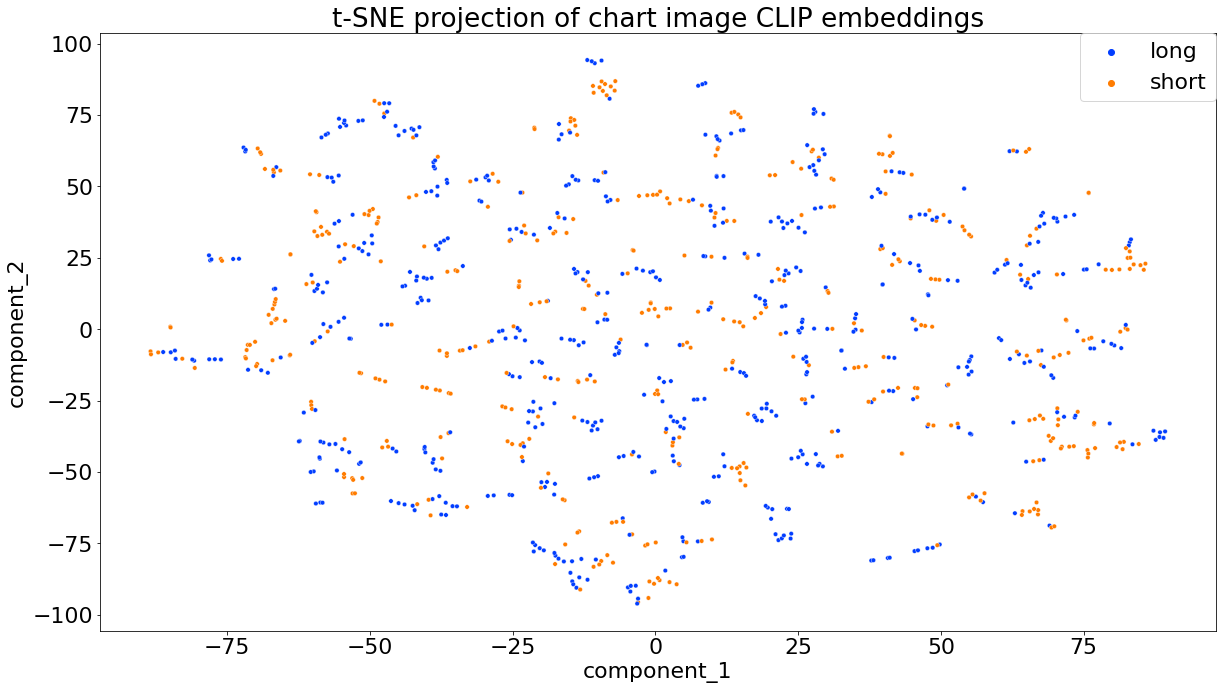}    \label{fig:thesis/images/0_6/TSNE_chart_images.png}}\quad
    \subfigure[CLIP-LSTM (text)]{\includegraphics[width=0.98\columnwidth]{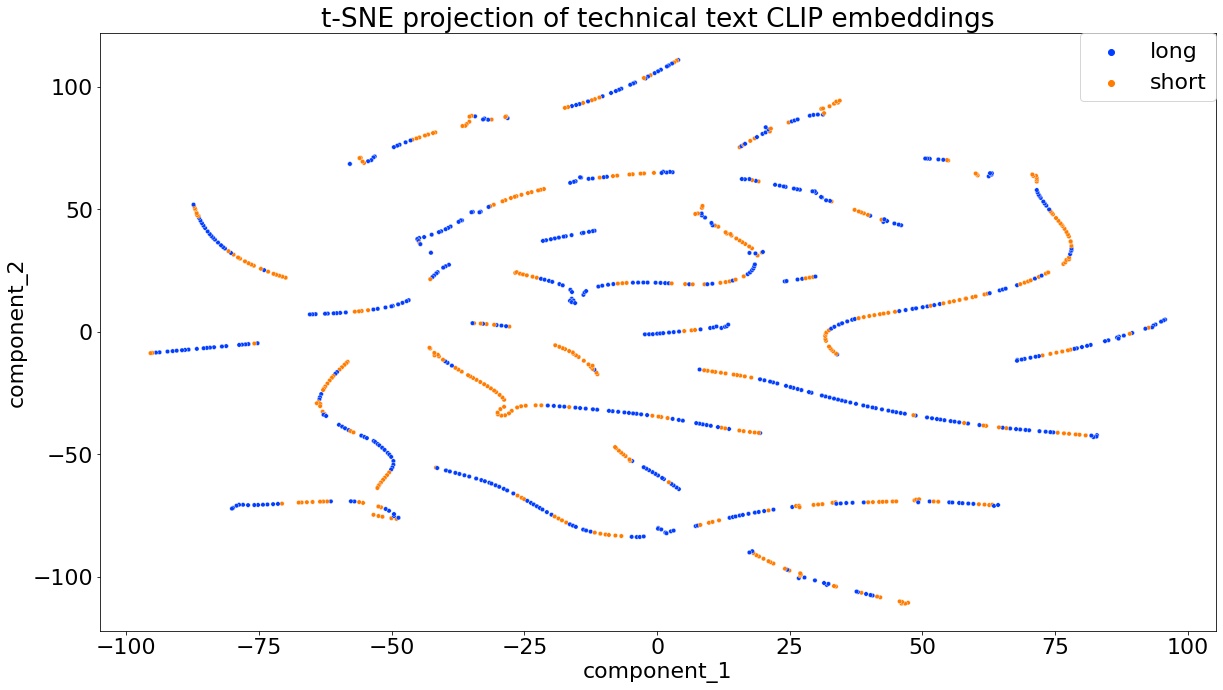}\label{fig:thesis/images/0_6/TSNE_technical_text.png}}
    \caption{Visualization of the feature vectors of the test set data points created with (a) CLIP-LSTM technical image and (b) CLIP-LSTM technical text. The embeddings are projected into two-dimensional space using t-SNE.}
\end{figure*}

\subsection{Evaluation Metrics} \label{Evaluation_experimentalDesign}
In the following, we explain the evaluation metrics for this study. We particularly aim to provide a wide range of evaluation metrics to cover different aspects of this prediction. In particular, due to the common existence of a high bias toward long positions (also in our test dataset), it is important to measure the success of a model also on short trades besides long trades. %\citet{asi4010013Text_Mining_of_Stocktwits_Data_for_Predicting_Stock_Prices} as an example, provide simply the F1 Score and accuracy. This, however, is not sufficient for an objective opinion, because with a strong long biased pricing, a network can only focus on long trades and the above-mentioned performance measurement possibilities would yield reliable statistics, but the real outcomes, both in the long and short term, are more informative. 

\paragraph{Precision} is an important metric for our experiments since it describes actual trading behavior when measured for each binary label. Precision is the ratio of the label's right hits to the total of the label's right hits plus the false positive specified hits, namely $\frac{\mathrm{TP}}{\mathrm{TP}+\mathrm{FP}}$. This metric indicates the correctness of trading activities in both directions when computed separately for the long and short label. We report precision for the long and short trades separately. 

%\paragraph{Recall} Recall is not used directly in this work, but it is still important for understanding in the subsequent course. Recall is the ratio of true positives to the total of true positives and false negatives.

% \begin{equation}
% \begin{aligned}
% \text {Precision} &=\frac{\mathrm{TP}}{\mathrm{TP}+\mathrm{FP}} \\
% \text {Recall} &=\frac{\mathrm{TP}}{\mathrm{TP}+\mathrm{FN}}
% \end{aligned}
% \end{equation}

\paragraph{F1 Score} is the harmonic mean of precision and recall, defined as $\frac{\mathrm{TP}}{\mathrm{TP} + 0.5 (\mathrm{FP} + \mathrm{FN})}$. We should however note that, as in this setting we are interested in the performance in both directions (long and short), F1 Score has the disadvantage that it is not using the True Negative (TN) values for calculation. In fact, F1 Score gives one value for a binary class problem and therefore pays more attention on the positive label (long trades). Despite this, F1 Score is commonly used in several previous studies~\cite{asi4010013Text_Mining_of_Stocktwits_Data_for_Predicting_Stock_Prices,Convolutionalneuralnetworkforstocktradingusingtechnicalindicators,8995193BERT_for_Stock_Market_Sentiment_Analysis}, and we will also report our results according to this metric.

\paragraph{Balanced Accuracy (Balanced ACC)} is particularly meaningful method in evaluating (binary) classifiers with unbalanced label distributions. This metric uses all four main parameters of the confusion matrix (TP, FP, TN, FN), and defined as:
\begin{equation}
\begin{aligned}
\text{Sensitivity}=\frac{\mathrm{TP}}{\mathrm{TP} + \mathrm{FN}}\quad  \text{Specificity}=\frac{\mathrm{TN}}{\mathrm{FP} + \mathrm{TN}}\\
\text{Balanced Accuracy}=\frac{\mathrm{Sensitivity} + \mathrm{Specificity}}{\mathrm{2}}
\end{aligned}
\end{equation}

\paragraph{Matthews Correlation Coefficient (MCC)} metric is insensitive to class imbalance and makes the assumption that the greater the correlation between true and predicted values are, the more accurate is the prediction for both labels. This metric is defined as:
\begin{equation}
\mathrm{MCC}=\frac{\mathrm{TP} \times \mathrm{TN}-\mathrm{FP} \times \mathrm{FN}}{\sqrt{(\mathrm{TP}+\mathrm{FP})(\mathrm{TP}+\mathrm{FN})(\mathrm{TN}+\mathrm{FP})(\mathrm{TN}+\mathrm{FN})}}
\end{equation}
The MCC metric becomes equal to $1$ if all samples are correctly categorized in the binary classes scenario, to $0$ if the classification is random, and to $-1$ if all samples are incorrectly classified.

\paragraph{Pip Balance} The aggregation of all completed trades is represented as the pip balance. This is the actual points received through trades and is a reliable indicator of the algorithm's actual performance.

\section{Results and Analyses}
\label{Results and Interpretability_MAIN}

In this section, we first cover the outcomes of all models in terms of the evaluation metrics separately for Standard and Delayed Labels. Next, we apply explainability methods to better understand the underlying functionalities of the CLIP-based models.

%The top three trained models for each strategy, in terms of the F1 Score in the validation set, are reported with the mean of the performances in the test set. 

%While F1 Score, MCC Score and Balanced ACC have a legitimate reason for analyzing a result, the literature (S. Kumar Chandar \cite{Convolutionalneuralnetworkforstocktradingusingtechnicalindicators} and Matheus Gomes de Sousa et al. \cite{8995193BERT_for_Stock_Market_Sentiment_Analysis} for example) demonstrates and evaluates the prediction precision for short trades much too rarely. This may result in misleading outcomes, since only algorithms with a high degree of precision in short trades really assess the price's true motion.

\subsection{Evaluation Results} 

\paragraph{Standard Label} 
Table \ref{tab:Scores_table_normal} (top) summarizes the evaluation results with the Standard Label approach. As reported, the CLIP-based models generally perform the best across the models over various evaluation metrics. As discussed before, due to the fact that stock markets are commonly biased towards positive direction in the long term, the precision of short transactions is critical.  Looking at the models' performance on short in Table~\ref{tab:Scores_table_normal}, we observe unequivocally the better performance of the CLIP text encoder, which is the only one to exceed the 50 percent short precision barrier. These results also indicate the assertion that the F1 Score is an inadequate metric for evaluating the performance of market prediction.

% Additionally, the test set's extremely positive price trend compounds this key attribute, since only those networks that have an abstracted comprehension of the real price trend also locate entry points in very positive price regions where the price swings downward over the following six hours.

%Furthermore, the F1 score in the Random CLIP-LSTM (Image) technique also provides an excellent value that is not represented in the Precision (Short) metric.

%This is a good illustration of the argument that the F1 score alone does not accurately reflect a network's trading productivity. To substantiate this assertion, Table~\ref{tab:results_pip_standard_technical} displays the actual traded pip throughout the test period.

\paragraph{Delayed Label} Table~\ref{tab:Scores_table_normal} (bottom) reports the evaluation results of the Delayed Label, namely the scenario where we wait one hour after prediction until the trade is executed. The label definition of Delayed Label implies that the technical information is already incorporated into the market movements. Similar to Standard Label, CLIP-based models show strong performance particularly in terms of Precision (Short) and Pip Balance (Short) to the pip, where only those networks that have a good score in the Precision (Short) metric also perform well in terms of total outcome. Given these results, contrary to expectations that waiting one hour after computing a trade's forecast would result in under performance across all model architectures, the achievement is on average highly comparable to the Standard Label approach model.

\subsection{Model Interpretation and Analysis} \label{Model Interpretation_results}

To better examine the underlying mechanism of the CLIP-based models, we apply out-of-the-box interpretation tools to the CLIP-LSTM (image) model. For interpretability reasons, the appropriate attention parameters are maintained per residual attention block during an image's forward pass, and backpropagation is performed with respect to the known output vector following the forward pass calculation. The computed information of all residual attention blocks is overlaid as a heatmap over the input picture after multiplying the two values (attention value and gradient) in the appropriate layers and subspaces. \footnote{Idea based on: https://github.com/hila-chefer} Figure~\ref{fig:thesis/images/0_6/true_long_interpret.png} depicts the chart images of three randomly sampled data points, in which the long direction was correctly predicted by the CLIP-LSTM (technical image) model. The images show the degrees of attentions of the CLIP model over various locations of the input chart image. The results show that the model is attending to various changes in the chart. Looking at these representative samples in this "True-Long" scenario, the model apparently distributes attention when the price is increasing, and puts more attention on the start and finish points when the stock price is decreasing.

We further examine the "True-Short" scenario (delta in six hours is negative and correctly predicted). We similarly randomly sample three "True-short" stock market input images shown in Figure~\ref{fig:thesis/images/0_6/true_short_interpret.png}. Looking at these results, an interesting finding is that the behavior inverts for the "True-short": when the chart is going downwards, the model places attention over whole the chart, and when the chart is ascending only the starting points are considered as important.

While so far we analyze the CLIP-LSTM (image) model, the question raises that how CLIP-LSTM (text), a model which is in principle designed for text/language processing, can also capture information about price and numbers. 
% To address this question, we should put emphasis on the nature of tokenization and the unique character of the CLIP network's training. 
Tokenization in CLIP is done at the byte level, enabling the text encoder component of CLIP to learn and recognize even numbers. To better show this point, Figure~\ref{fig:thesis/images/0_6/clip_text_interpret.png} in Appendix~\ref{sec:appendix:experiment} displays the embeddings of the numbers from $1$ to $1000$, each given as a string to the CLIP text encoder, and projected into 2-dimensions for the sake of visualization. The figure demonstrates that the CLIP model clusters the numbers in similar ranges in closer proximities and hence is aware of the magnitude of the numbers.

We further demonstrate the feature vectors created by the CLIP-based models. Figure~\ref{fig:thesis/images/0_6/TSNE_chart_images.png} illustrates the embeddings of the technical images generated with CLIP, projected into two dimensions using the t-SNE method. As shown, the model can well separate and cluster the data points with various labels. Figure~\ref{fig:thesis/images/0_6/TSNE_technical_text.png} shows the same method on the data points using the embeddings of the technical text data. In the case of the text data, CLIP can clearly detect a temporal sequence shown as distinct trajectories in the two-dimensional space. 
% Also in this plot, we observe the frequent appearance of identical labels in sequences, suggesting not only relations over time but also grouping according to specific labels. 

%In Appendix C (supplementary materials), we provide further investigations on the characteristics of CLIP in learning representations for numerical data when provided in the form of a text sequence.

\section{Conclusion}
\label{Conclusion_MAIN}
In this study, we conduct a large set of experiments on forecasting the German share index using various deep learning approaches. We introduce two novel forecasting models based on the pre-trained CLIP vision-language model. These models predict market movements by processing text and image data, provided by transforming the original stock market data into sequences of texts and images of price charts, respectively. Our findings show that the CLIP models both with the image and text encoder, significantly outperform strong baseline models. Our further experiments show that our models does not necessarily bounded on a quick reactions, and can still be profitable when the prediction is executed with one hour delay.

%\vspace{-2mm}
\section{Limitations} 
\label{sec:limitations}
A limitation of this work is the scope and size of the dataset, particularly when considering the stock market's acquisition and various influential components in the market. Considering this, a specific behavior for a specified time period might be especially profitably, but not necessarily generalizable to other unseen future cases. To mitigate this problem, we train the networks over a highly volatile period characterized by big stock market downturns and subsequent dramatic stock market increases. Despite this, a future step would be to examine the proposed method on the data in other markets and data periods. Additionally, computational resources particularly needed for the CLIP-based approaches is a constraint of out proposed approach.

%%%%%%%%%%%

%\vspace{-2mm}
\bibliography{literature}

\appendix

\section{Technical Terminologies in Financial Domain} \label{sec:appendix:background-finance}

% The most significant stock market words are given here in order to be able to use technical phrases in the next sections of the thesis and to clarify the main goal. These are essential for general understanding and have been condensed to the most important keywords.
The most important stock market terms are included here in order to facilitate the usage of technical terms in subsequent parts of the thesis and to define the primary objective.

\begin{figure}[tb!]
\centering
    \subfigure[]{\includegraphics[width=0.92\columnwidth]{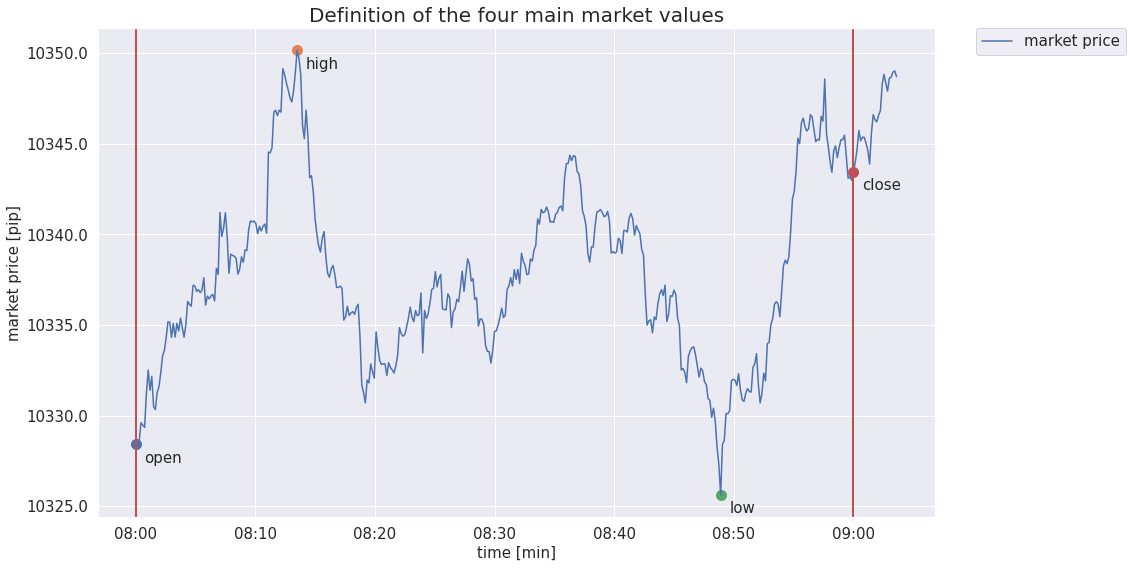}\label{fig:thesis/images/0_2/HighLowOpenClose.png}
    }
    
    \subfigure[]{\includegraphics[width=0.92\columnwidth]{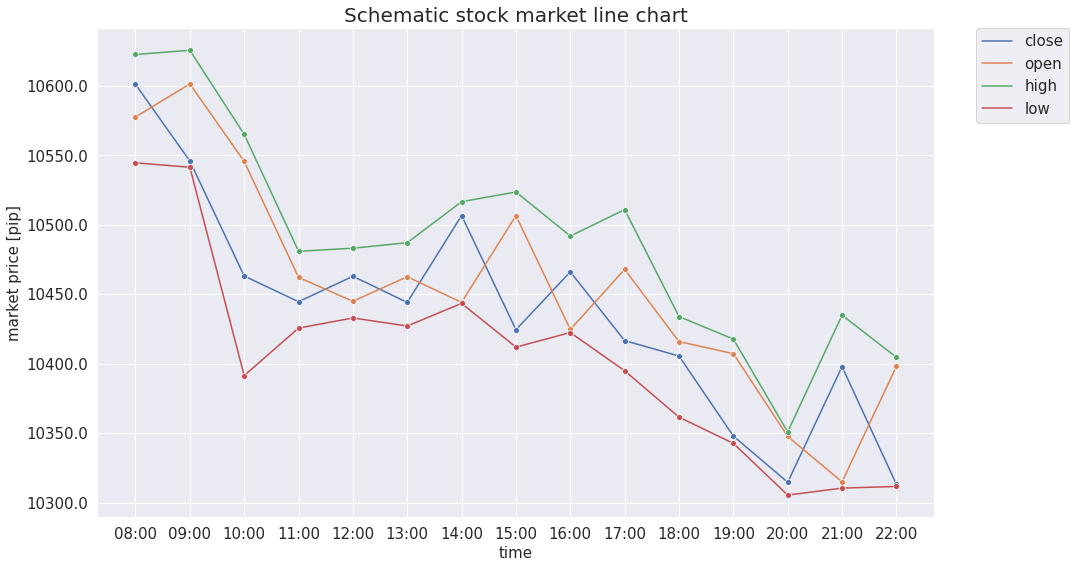}\label{fig:thesis/images/0_2/LineChart.png}}

    \subfigure[]{\includegraphics[width=0.92\columnwidth]{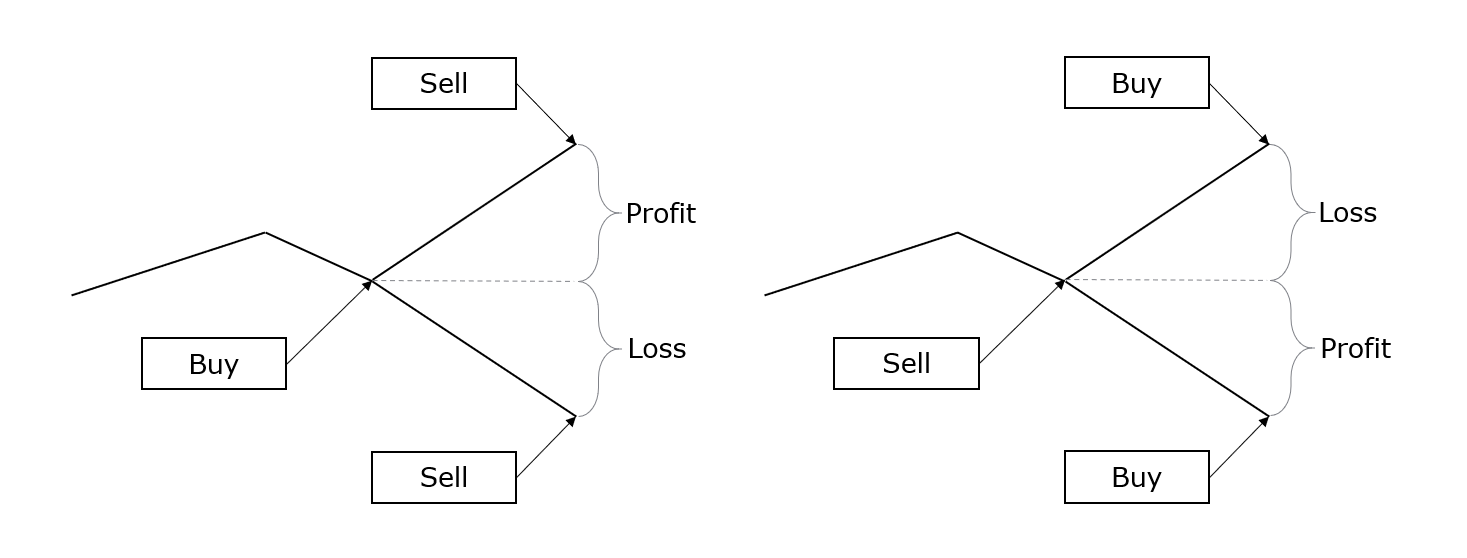}\label{fig:thesis/images/0_2/Long_short.png}}
\caption{(a) Schematic explanation of the composition of the four main values in the stock market using the underlying continuous market movement defined on the one-hour interval. (b) Schematic representation of a line chart on an hourly sample time of the stock market. (c) Schematic explanation of the buy (long) and sell (short) trading options in Contract for Difference (CFD).}
\end{figure}

\paragraph{Technical Data} 
A share's technical data, in general, does not include any information on macroeconomic or company-specific statistics. It is solely defined by price movement, and it is based on the scanning of the continuous course of the stock. For a one-hour sampling rate, technical data of a stock consists of four separate values, made available by various suppliers for at least one year into the past. These four numbers are the highest price in the designated hour, the lowest price, the opening price, and the closing price. Figure~\ref{fig:thesis/images/0_2/HighLowOpenClose.png} depicts of the four major values during the hour from 08:00 to 09:00 of a fictitious share price.

\paragraph{Intraday Trading}
Buying and selling a stock in a short period of time is an increasingly common financial strategy, particularly among private investors. Many services offer buy and sell ads down to the smallest amount of money. The usage of the hourly chart, namely displaying the course with one scan each hour, is highly popular and is regarded to be essential information in the subsequent course.

\paragraph{Line Chart} 
The line chart is defined by the points of the four major values connected in a certain time interval pattern. The four primary values are therefore acquired for each hour, forming four lines plotted over time. Figure~\ref{fig:thesis/images/0_2/LineChart.png} depicts an exemplary line chart with a one-hour sampling basis.

% \begin{figure*}[t]
%     \centering
%     \includegraphics[width=0.8\textwidth]{thesis/images/0_2/DAIN.png}
%     \caption{Schematic representation of the Deep Adaptive Input Normalization (DAIN) approach workflow. Image from \citet{passalis2019deepDeep_Adaptive_Input_Normalization_for_Time_Series_Forecasting}}
%     \label{fig:thesis/images/0_2/DAIN.png}
% \end{figure*}

% \begin{figure*}[t]
%     \centering
%     \includegraphics[width=0.8\textwidth]{thesis/images/0_2/CLIP_architekture.png}
%     \caption{Overview for the CLIP approach. CLIP trains the right pairing of the generated image and text embeddings. Image from~\citet{CLIP}}
%     \label{fig:thesis/images/0_2/CLIP_architekture.png}
% \end{figure*}

\paragraph{Long and Short Trading}
Contract for Difference (CFD) trading allows not only buying a stock as normal, but also to initiate a short transaction. As a consequence, if the algorithm decides to open a long trade, the deposit would be positive if the price of the stock rises in the future, and vice versa, if the algorithm opens a short trade, the deposit will be positive if the chart drops below the trade's entry level. To summarize: Long Trade is positive if stock chart moves up in future and is negative if stock chart moves down in future. Short Trade is positive if stock chart moves down in future and is negative if stock chart moves up in future. Figure \ref{fig:thesis/images/0_2/Long_short.png} summarizes the two possible Contract for Difference (CFD) trading options.

\paragraph{Pip}
A stock's movement in the stock market is measured in "pip", the point value at which the market is standing at a certain point in time. For example, if a long trade begins at 14000 pip and the stock climbs to 14050 after a variable period of time, the net profit -- assuming that the trade is concluded at this level -- is 50 pip.

\section{Background on Deep Learning Architectures} 
\label{sec:appendix:background-deep} 
This part brings together established deep-learning architectural principles and novel neural network architectures for the purpose of market prediction, which this thesis investigates and analyzes.

\paragraph{LSTM} \label{LSTM_background}
Conventional Recurrent Neural Networks (RNN) are incapable of using information from the distant past. Hochreiter and Schmidhuber approached this limitation of RNNs by introducing the Long Short Term Memory \cite{hochreiter1997longLong_short_term_memory} (LSTM), that is capable of memorizing data across plenty of intervals. It does this by retrieving information from the past and storing it in what is known as the cell state. This cell state is managed each time step by the forget gate (which forgets some information) and is supplemented by the input gate's outputs.

\paragraph{DAIN} 
Deep Adaptive Input Normalization \cite{passalis2019deepDeep_Adaptive_Input_Normalization_for_Time_Series_Forecasting} (DAIN) is a deep-learning data normalization approach, which separately learns the normalizing parameters for a given data set. The first stage is responsible for relocating the data inside the feature space, a process known as centering. The data is then linearly scaled in the second stage to affect the variance of the data. This is also referred to as standardizing. The final stage is to retain attributes in a nonlinear fashion for those that are irrelevant or inappropriate, which is called gating. The DAIN approach is adaptive, because of the normalization scheme that is used is determined by the actual time series supplied into the network, and trainable, in that the suggested layer's behavior is tailored to the job at hand via back-propagation. %Figure~\ref{fig:thesis/images/0_2/DAIN.png} illustrates the DAIN approach.

\begin{table*}[t]
\centering
\begin{tabular}{lcccc}
\toprule
                              \textbf{Model} & \textbf{bs} & \textbf{lr} & \textbf{drop} & \textbf{w} \\
\midrule
                                LSTM &   128 & 0.001 &  0.4 & -0.00005 \\
                              LSTM* &   128   & 0.001 &  0.4 & -0.00005 \\
                LSTM (long sequence) &   128 & 0.001 &  0.4 & -0.00005 \\
              LSTM (long sequence)* &   64 & 0.001 &  0.2 & -0.00005 \\
        Stacked-LSTM (long sequence) &   16 & 0.001 &  0.4 & -0.00005 \\
      Stacked-LSTM (long sequence)* &   16 & 0.001 &  0.4 & -0.00005 \\
                          DAIN-LSTM &   256 & 0.0005 &  0.2 & -0.00015 \\
                          DAIN-LSTM* &   256 & 0.0005 &  0.2 & 0.0 \\
            Random CLIP-LSTM (image) &   32 & 0.00005 &  0.2 & 0.0 \\
          Random CLIP-LSTM (image)* &  32 & 0.0005 &  0.4 & -0.00015 \\
                  CLIP-LSTM (image) &   128 & 0.00005 &  0.4 & 0.0 \\
                  CLIP-LSTM (image)* &   128 & 0.00005 &  0.4 & 0.0 \\
             Random CLIP-LSTM (text) &   64 & 0.00005 &  0.4 & -0.00015 \\
            Random CLIP-LSTM (text)* &   32 & 0.00005 &  0.0 & -0.00015 \\
                    CLIP-LSTM (text) &   64 & 0.00005 &  0.0 & -0.00015 \\
                  CLIP-LSTM (text)* &   128 & 0.0005 &  0.2 & -0.00015 \\
          CLIP-LSTM (image and text) &   64  & 0.00005 &  0.0 & 0.0 \\
         CLIP-LSTM (image and text)* &   128 & 0.0001 &  0.4 & 0.0 \\
                      
\bottomrule
\end{tabular}
 \caption{Hyperparameters of the best performing models on F1 Score validation set basis, where models with * denote the Delayed Labels and without * denote the Standard Labels. Due to space limitations, "image" is used for "technical image", "text" is used for "technical text" and "image and text" is uses for "technical image and technical text".}
\label{hyperparameter_tab}
\end{table*}

% \begin{table*}[t]
% \centering
% \begin{tabular}{lccccccc}
% \toprule
%         \textbf{Model} &     \textbf{F1} &     \textbf{MCC} &  \textbf{ACC} & \textbf{Precision}     & \textbf{Precision} &       \textbf{Balance} &       \textbf{Balance}\\
%                        &     \textbf{Score} &  \textbf{Score}     &  \textbf{Balanced}      & \textbf{(Short)} & \textbf{(Long)} &       \textbf{Short} &       \textbf{Long}\\
% \midrule
% Random                          & 0.52  & -0.09                      & 0.45  &   36.91  &  54.01  & -1049.19                            & -115.69  \\
% Always long                     & 0.73  &  0.00  & 0.50  &   0.00  &  58.03  & 0.00        &  933.49 \\
% Always short                    & 0.00  &  0.00  & 0.50  &   41.96  &  0.00  &  -933.49         &  0.00 \\
% \bottomrule
% \end{tabular}
% \caption{I don't know exactly.}
% \end{table*}

\paragraph{CLIP} \cite{CLIP} is a OpenAI-developed and trained state-of-the-art approach, that trains a text encoder as well as an image encoder using two multi-layer transformer \cite{AttentionIsAllYouNeed} architectures. 
This architecture's unique kind of training enables the network to acquire visual concepts via natural language supervision because of the way it was created. The purpose of this technique is to find the right text-image pairings in a batch given that occur on the dot product matrix's major diagonal, and so the network has a known objective of raising the main diagonal's values while decreasing all other values. The model is trained on 400 million picture-text pairings in order to accomplish the objective of simultaneously comprehending the text and the content of the corresponding image. CLIP \cite{CLIP} outperforms a number of fully supervised trained models and shows strong performance on zero shot challenges.

\begin{figure*}[t!]
    \centering
    \includegraphics[width=0.8\textwidth]{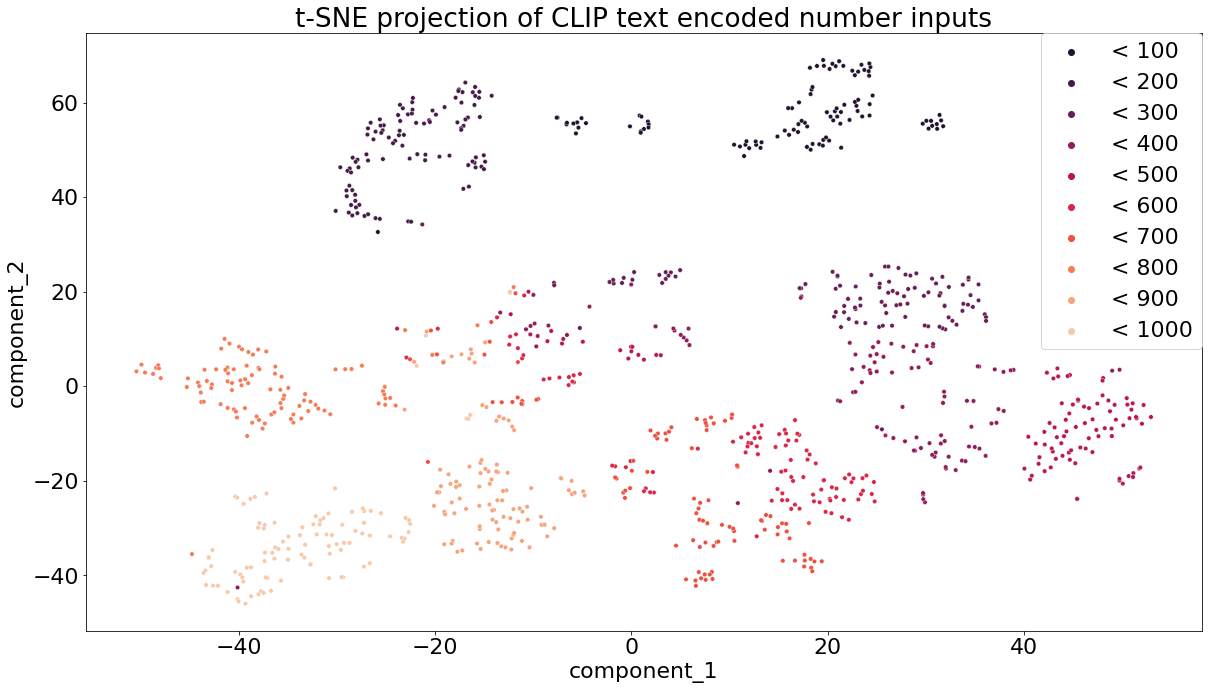}
    \caption{The embeddings of the numbers given to the CLIP text encoder using t-SNE projection. The numbers categorized into ranges form clusters showing that CLIP provides meaningful embeddings in respect to the magnitudes of the numbers.}
    \label{fig:thesis/images/0_6/clip_text_interpret.png}
\end{figure*}

The novel strategy of training gives CLIP an outstanding abstraction comprehension as it was not taught on classes but on written texts with the corresponding images as input. Thus, it is compelled to concentrate its attention on the substance of the words and the complete content of the picture. 
A significant distinction between the CLIP text encoder and other language models is the byte-level tokenization. This provides enormous benefits in terms of performance and, more specifically, the understanding of numbers as language. Figure~\ref{fig:thesis/images/0_6/clip_text_interpret.png} shows increasing numerical values that are encoded and projected using t-SNE on two planes to demonstrate the CLIP text encoder's perception of numerical values. This example demonstrates that the CLIP text encoder can accurately discriminate between numbers of varying sizes in high-dimensional space.

\section{Additional Experiment Setup} \label{sec:appendix:experiment}

The hyperparameters of the models are reported in Table~\ref{hyperparameter_tab}. The batch size is denoted by bs, lr is the learning rate, drop denotes dropout and w denotes the weighting of the cross entropy loss in this expression. This parameter, which we use to raise or reduce the effect of short samples during the training of the network, is useful since the two classes, short and long, do not have the same number of samples. It was necessary to apply three additional learning rates for the DAIN-LSTM network for the three steps of training this model, initialized in the same manner as stated in \cite{passalis2019deepDeep_Adaptive_Input_Normalization_for_Time_Series_Forecasting}. %Table~\ref{all_results_pip} shows that each model has the same standard deviations for both short and long balance. This is because the network's long and short transactions are directly reliant on the earnings and losses on the opposite side. More precisely, if in the test set a long transaction would be replaced by a short transaction, the long depots falls, while the short depots drops as well, based on the network's preference in each scenario. The ensuing outcomes move the mean value and modify the standard deviation value, while the standard deviation values per network has the same value.

%\paragraph{Why do CLIP-based models perform well?} To address this question, we should put emphasis on the nature of tokenization and the unique character of the CLIP network's training. Tokenization in CLIP is done at the byte level, enabling the text encoder component of CLIP to learn and recognize even numbers. To better show this point, Figure~\ref{fig:thesis/images/0_6/clip_text_interpret.png} displays the embeddings of the numbers from $1$ to $1000$, each given as a string to the CLIP text encoder, and projected into 2-dimensions for the sake of visualization. The figure demonstrates that the CLIP model clusters the numbers in similar ranges in closer proximities and hence is aware of the magnitude of the numbers.

\end{document}